\documentclass[
superscriptaddress,
twocolumn,
amsmath,amssymb,
aps,
pra,
]{revtex4-2}

\usepackage{graphicx,framed} 
\usepackage{dcolumn} 
\usepackage{bm} 
\usepackage{braket}
\usepackage{xcolor}
\usepackage{amsthm}
\usepackage{hyperref} 
\usepackage[normalem]{ulem} 

\usepackage{amsthm}

\usepackage{comment}

\usepackage{float}
\makeatletter
\let\newfloat\newfloat@ltx
\makeatother
\usepackage{algorithm}
\usepackage[noend]{algpseudocode}
\algdef{SE}[DOWHILE]{Do}{doWhile}{\algorithmicdo}[1]{\algorithmicwhile\ #1}
\makeatletter
\def\algbackskip{\hskip-\ALG@thistlm}
\makeatother

\definecolor{lightblue}{RGB}{73,151,208}
\definecolor{crimson}{RGB}{140,41,53}

\hypersetup{
    colorlinks,
    linkcolor={crimson},
    citecolor={lightblue},
    urlcolor={lightblue}
}

\newcommand{\gammath}{\gamma_R}
\newcommand{\gammain}{\gamma_\mathrm{in}}
\newcommand{\gammaout}{\gamma_\mathrm{out}}

\newcommand{\mN}{\mathcal{N}}
\newcommand{\mD}{\mathcal{D}}

\newcommand{\gin}{g_{\mathrm{in}}}
\newcommand{\gout}{g_{\mathrm{out}}}

\newcommand{\goutb}{g_{\mathrm{out}}^{\beta}}
\newcommand{\ginb}{g_{\mathrm{in}}^{\beta}}

\newcommand{\goutbbar}{\overline{g}_{\mathrm{out}}^{\beta}}
\newcommand{\ginbbar}{\overline{g}_{\mathrm{in}}^{\beta}}

\begin{document}

\preprint{}
\title{Genuinely Out-of-Equilibrium Bose–Einstein Condensation}
\date{\today}
\author{Tatsuhiko N. Ikeda}
\affiliation{Faculty of Social Informatics, ZEN University, Zushi, Kanagawa, 249-0007, Japan}
\affiliation{RIKEN Center for Quantum Computing, Wako, Saitama 351-0198, Japan}
\affiliation{Department of Applied Physics, Hokkaido University, Sapporo, Hokkaido, 060-8628, Japan}

\author{Masahito Ueda}
\affiliation{%
Department of Physics, University of Tokyo, 7-3-1 Hongo, Bunkyo-ku, Tokyo, 113-0033, Japan}
\affiliation{
 Institute for Physics of Intelligence, University of Tokyo, 7-3-1 Hongo, Bunkyo-ku, Tokyo, 113-0033, Japan}
 \affiliation{Fundamental Quantum Science Program (FQSP), TRIP Headquarters, and RIKEN Center for Emergent Matter Science (CEMS), RIKEN, Wako 351-0198, Japan}%

\begin{abstract}
Bose-Einstein (BE) condensation typically occurs in the ground state.  Prominent exceptions include excitons, whose ground state is the vacuum; consequently, they undergo BE condensation only in out-of-equilibrium regimes to counteract rapid decay. Here, we develop a driven-dissipative framework to address this challenge, demonstrating that BE condensates emerge under sufficiently strong excitation in one and two dimensions. 
In three dimensions, by incorporating effective loss rates and a specified injection energy, we find a percent-level condensate fraction, consistent in order of magnitude with the exciton experiment of Morita et al. (Nat. Commun. {\bf 13}, 5388 (2022)).
\end{abstract}

\maketitle

\section{Introduction}

Bose-Einstein (BE) condensation of ideal or weakly interacting bosons occurs when the thermal de Broglie wavelength exceeds the average interparticle distance \cite{Pitaevskii2016, Leggett2001}. This phenomenon was rigorously established by Einstein for ideal bosons \cite{Einstein1925} and by Kennedy et al. for a zero-temperature lattice gas at half-filling \cite{Kennedy1988}. Furthermore, Gavoret and Nozi\`{e}res demonstrated that a  BE condensate survives when  interactions are switched on adiabatically within a perturbative regime \cite{Gavoret1964}.
While the existence of BE condensation has been widely verified experimentally for weakly interacting bosons \cite{Anderson1995, Davis1995, Bradley1995}, it is not universally guaranteed for strongly interacting systems, as evidenced by the crystalline-solid phase of ${}^4$He. An even more challenging open question is whether BE condensates can emerge in highly dissipative systems, such as excitons, whose ground state is the vacuum.

Traditionally, BE condensation is formulated within the framework of equilibrium statistical mechanics \cite{Dalfovo1999}. In contrast, a vast array of contemporary physical systems--ranging from exciton-polaritons in semiconductor microcavities \cite{Kasprzak2006, Deng2002, Byrnes2014} and photons in dye-filled cavities \cite{Klaers2010} to magnons \cite{Demokritov2006} and excitons in bulk semiconductors \cite{Snoke2002, Lin1993}--are inherently open systems driven by a continuous influx of energy or particles while being subject to unavoidable decay. These driven-dissipative systems operate far from thermal equilibrium, eventually settling into non-equilibrium steady states, where the intricate interplay between gain, loss, interaction, and thermalization governs their dynamic behavior \cite{Carusotto2013, Sieberer2013}. Understanding BE condensation in such far-from-equilibrium situations not only presents fundamental theoretical challenges \cite{Szymanska2006, Wouters2007} but also provides profound insights into driven-dissipative quantum condensation.

The study of excitonic BE condensation has a long and rich history, motivated by the prospect of achieving condensation at relatively high temperatures due to the light effective mass of excitons \cite{Keldysh1968, Combescot2008}. Despite decades of intense research, the unambiguous observation and theoretical understanding of this phenomenon--particularly in bulk materials like Cu$_2$O--have been hampered by several intrinsic challenges. These include a finite exciton lifetime caused by radiative and non-radiative recombination, intricate thermalization pathways, and interactions that can cause deviations from ideal Bose gas behavior or trigger transitions into other phases such as electron-hole plasmas and biexcitons \cite{Snoke2014}. As highlighted in pioneering theoretical works and subsequent experimental endeavors \cite{Kavoulakis1996}, understanding and mitigating loss mechanisms to achieve sufficient phase-space density and cooling remains a critical hurdle. Their work underscored the complex dynamics, including Auger recombination and biexciton formation, that compete with and potentially obscure condensate formation, especially under nonequilibrium conditions where continuous pumping is required.

A fundamental question posed by Kohn and Sherrington \cite{Kohn1970} centers on the very nature of an exciton condensate: whether it represents a true BE condensation decoupled from the radiation field, or a more complex coherent state intrinsically linked to light, the excitonic insulator concept also being relevant here \cite{Keldysh1965}. This inquiry is particularly pertinent in genuinely out-of-equilibrium systems, where the condensate phase is fundamentally shaped by the balance of gain and loss. Resolving these open questions requires theoretical frameworks that transcend equilibrium paradigms by explicitly incorporating the dynamics of particle creation, annihilation, dissipation and thermalization in open quantum systems. While kinetic equations and Keldysh techniques have yielded valuable insights \cite{Keeling2004}, a master equation approach \cite{Breuer2002, Gardiner2004, Carmichael1999} provides a powerful, versatile tool for describing the time evolution of the system's density matrix, naturally unifying coherent dynamics of quantum condensation with incoherent thermalization processes arising from system-bath couplings.

A major experimental breakthrough was recently achieved by Morita et al. \cite{Morita2022}, who reported the direct observation of a 1s paraexciton  condensate in bulk Cu$_2$O under far-from-equilibrium conditions. Their experiments, utilizing mid-infrared induced absorption imaging, provided compelling evidence for a spatially localized condensate forming in a trapped gas of excitons in contact with a cold phonon bath and subject to the inherent particle losses. This landmark achievement establishes a crucial experimental benchmark, calling for a comprehensive theoretical framework that captures the essential physics of genuinely out-of-equilibrium BE condensation--particularly the behavior of the condensate fraction under varying regimes of particle gain and loss. The foundational role of such nonequilibrium regimes was also underscored early on in the context of polariton lasing \cite{Imamoglu1996}.

In this paper, we develop a driven-dissipative approach to BE condensation based on a configuration-space master equation that accounts for particle gain and loss, as well as the thermalization process leading to condensation. This framework enables the evaluation of key observables, in particular the condensate fraction, thereby providing a quantitative measure of condensation in nonequilibrium systems. Our work is inherently motivated by the intrinsic non-equilibrium nature of exciton condensates, whose ground state corresponds to the vacuum \cite{Morita2022}. The developed master equation naturally incorporates particle injection and loss, enabling a systematic investigation into how these rates, in tandem with thermal effects, dictate the properties of the steady-state condensate.

Focusing on the intricate interplay of coherent evolution, environmental coupling, particle gain and loss, and thermalization, this work elucidates the fundamental features of far-from-equilibrium BE condensation. We show that BE condensates can emerge under sufficiently strong excitation in one and two dimensions. This represents a striking departure from the Hohenberg-Mermin-Wagner theorem, which precludes thermal-equilibrium low-dimensional BE condensation at finite temperatures \cite{Hohenberg1967,MerminWagner1966,Mermin1968}.
Furthermore, our theoretical approach enables a direct comparison with the experimental findings in Cu$_2$O. Incorporating discrete energy levels, effective loss, injection energy, and a phenomenological relaxation coefficient, we find a condensate fraction consistent in order-of-magnitude agreement with the observed percent-level value. 
This framework provides crucial insights into the stability and thermodynamics  of open dissipative condensates, paving the way for further exploration into the burgeoning field of non-equilibrium BE condensation.

\section{Configuration-space master equation}
\label{sec:Configuration-space master equation}

We consider a dilute, non-interacting Bose gas subject to 
(i) energy exchange with a heat bath, 
(ii) laser pumping (particle inflow), and 
(iii) particle loss via trap imperfections and other decay mechanisms (particle outflow).  
The single-particle eigenstates are indexed by
$(k,\alpha)$, where $k=0,1,\dots$ labels the energy $\epsilon_k$, and
$\alpha=1,\dots,M_k$ distinguishes the $M_k$ degenerate sublevels.
Let $a_{k\alpha}^\dag$ and $a_{k\alpha}$ be the creation and annihilation operators of a particle in state $(k,\alpha)$. The Hamiltonian then reads
\begin{equation}
  H=\sum_{k,\alpha}\epsilon_k\,a_{k\alpha}^\dag a_{k\alpha}.
\label{Hamiltonian}
\end{equation}
A many-body Fock state is defined as 
\begin{equation}
\ket{\vec N}:=\ket{\{N_{k\alpha}\}}=
 \prod_{k,\alpha} \frac{1}{\sqrt{N_{k\alpha}!}}(a_{k\alpha}^\dag)^{N_{k\alpha}}\ket{0}.
\label{configuration}
\end{equation}
This state is a simultaneous eigenstate of the Hamiltonian and the particle-number operator
\begin{equation}
\hat{n}_{k\alpha}:=a_{k\alpha}^\dagger a_{k\alpha}
\label{particle-number} 
\end{equation}
for all $(k,\alpha)$, and will serve as a basis state in the following discussions.

\subsection{GKSL description: microscopic processes}

The density operator $\rho$ of the system is assumed to obey the Gorini–Kossakowski–Sudarshan–Lindblad (GKSL) equation
\begin{equation}
  \dot\rho=-i[H,\rho]+\mD_{\rm pump}[\rho]+\mD_{\rm loss}[\rho]+\mD_{\rm R}[\rho],
 \label{eq:GKSL}
\end{equation}
where the first term on the right-hand side accounts for the unitary evolution under the Hamiltonian (\ref{Hamiltonian}), while the others represent dissipators that describe the nonunitary dynamics caused by the laser pump, particle loss, and thermal relaxation. Each dissipator is assumed to take the canonical form
\begin{equation}
\mD[\rho]=\sum_\mu(L_\mu\rho L_\mu^\dag-\tfrac12\{L_\mu^\dag
L_\mu,\rho\}),
\label{dissipator}
\end{equation}
which ensures a completely positive trace-preserving dynamics specified by the jump operators $L_\mu$.

\subsubsection{Pump: $\mathcal{D}_{\rm pump}$}
We consider a situation in which bosons are excited by a laser field that populates a designated level $k_G$.  The corresponding jump operators are given by
\begin{equation}
  L_{{\rm up},k_G,\alpha}^{\rm pump}=\sqrt{\gamma_{\rm in}}\;a_{k_G\alpha}^\dag, \ \ \ 
  L_{{\rm down},k_G,\alpha}^{\rm pump}=\sqrt{\gamma_{\rm in}}\;a_{k_G\alpha},
  \label{jump_gain}
\end{equation}
where the former creates a boson at level $k_G$ via photon absorption, while the latter annihilates one through stimulated emission. 

\subsubsection{Loss: $\mathcal{D}_{\rm loss}$}
The loss term accounts for the effects of imperfections, such as boson leakage  from the trapping potential and spontaneous decay into dissociated exciton modes.  These loss processes remove a boson from any state $(k,\alpha)$ via  the jump operators
\begin{equation}
  L_{k\alpha}^{\rm loss}=\sqrt{\gamma_{\rm out}}\;a_{k\alpha}.
\label{jump_loss}
\end{equation}
Crucially, the pump affects only the designated energy level $k_G$,  whereas the loss occurs at all levels since there is no a priori reason to assume any spectral selectivity. Any energy dependence of the loss coefficient $\gamma_{\rm out}$ arising from experimental realities is neglected here for  simplicity.

\subsubsection{Thermal relaxation: $\mathcal{D}_{\rm R}$}
Thermal relaxation proceeds via energy exchange with a heat bath (e.g.\ phonons), causing bosons to undergo transitions between energy levels.  
The associated jump operators read 
\begin{align}
  L_{k\alpha\to k'\alpha'}
  &=\sqrt{\Gamma_{k\alpha\to k'\alpha'}}\,
    a_{k'\alpha'}^\dagger a_{k\alpha}.
\label{thermal_jump}
\end{align}
Assuming the system is in contact with a heat bath at inverse temperature $\beta$, the transition rates (\ref{thermal_jump}) and their time-reversal counterparts $L_{k'\alpha'\to k\alpha}$ fulfill the Kubo–Martin–Schwinger (KMS) condition \cite{Kubo1957,Martin1959,Breuer2002}:
\begin{equation}
e^{-\beta\epsilon_k}\Gamma_{k\alpha\to k'\alpha'}
 =e^{-\beta\epsilon_{k'}}\Gamma_{k'\alpha'\to k\alpha},  \label{eq:KMS}
\end{equation}
which guarantees detailed balance in the absence of pump and loss.

\subsection{Transforming the density matrix to the configuration-space probability distribution}

Equation~\eqref{eq:GKSL} completely determines the quantum dynamics of the system.  In dilute gases of excitons, phase coherence between different Fock states typically decays much faster than the rate of change in occupation numbers; hence, retaining only the diagonal matrix elements provides a quantitatively accurate description.

Even after restricting the analysis to diagonal elements, solving Eq.~(\ref{eq:GKSL}) to investigate the BE transition presents a formidable challenge due to the interplay of competing physical processes and the associated vast dimension of the Hilbert space.
To overcome this difficulty, we map the density matrix onto a probability distribution in configuration space by projecting it onto the Fock basis:
\begin{equation}
P(\vec N)=\braket{\vec N|\rho|\vec N},
\end{equation}
where ${\vec N}=\{N_{k,\alpha}\}$ denotes the full set of occupation numbers. Under this diagonal approximation, 
Eq.~\eqref{eq:GKSL} reduces to a classical master equation
\begin{equation}
  \dot P(\vec N)=
  D_{\rm pump}[P]+D_{\rm loss}[P]+D_{\rm R}[P],
  \label{eq:MasterP}
\end{equation}
where the dot denotes the time derivative and $D_\chi[P]$ ($\chi=$pump, loss, R) represents the contribution from each dissipator within the diagonal approximation.

Equations~\eqref{eq:GKSL}–\eqref{eq:MasterP} provide a systematic framework for analyzing driven-dissipative dynamics. While the GKSL equation captures quantum coherence when necessary, its diagonal projection yields a
computationally efficient kinetic equation that incorporates both the
key bosonic enhancement factor responsible for BE condensation and the detailed-balance structure that ensures thermal contact with heat reservoirs such as a substrate and a coolant.

We now discuss the concrete form of each dissipator.

\subsubsection{Pump}
The pump process is governed by the two jump operators given in Eq. (\ref{jump_gain}), and thus consists of two distinct contributions:
\begin{equation}
D_{\rm pump}[P]=D_{\rm pump}^{\rm up}[P]+D_{\rm pump}^{\rm down}[P].
\end{equation}
Here $D_{\rm pump}^{\rm up}[P]$ describes the creation of a boson in state $(k_G,\alpha)$ via laser photon absorption:
\begin{eqnarray}
D_{\rm pump}^{\rm up}[P(\vec{N})]&=&\gamma_{\rm in}\sum_\alpha \left[N_{k_G\alpha}P(\vec{N}-\vec{e}_{k_G\alpha})
\right. \nonumber \\
& & \left. \hspace{0.9cm}
-(N_{k_G\alpha}+1)P(\vec{N})
\right],
\label{gain_A}
\end{eqnarray}
where the first term on the right-hand side represents a transition from  a   configuration containing $N_{k_G\alpha}-1$ bosons in the $(k_G,\alpha)$ mode. The addition of a particle to this mode then drives the system into the target configuration $\vec{N}$. The second term describes a process originating from the configuration $\vec{N}$, so that a particle gain in the same mode decreases the probability $P(\vec{N})$.

On the other hand, $D_{\rm pump}^{\rm down}[P]$ describes the  annihilation of a boson from state $(k_G,\alpha)$ via stimulated photon emission:
\begin{eqnarray}
D_{\rm pump}^{\rm down}[P(\vec{N})]&=&\gamma_{\rm in}\sum_\alpha \left[(N_{k_G\alpha}+1)P(\vec{N}+\vec{e}_{k_G\alpha})
\right. \nonumber \\
& & \left. \hspace{1cm}
-N_{k_G\alpha}P(\vec{N})
\right].
\label{gain_B}
\end{eqnarray}
The differences in the coefficients of $P$ between Eq.~(\ref{gain_A}) and Eq.~(\ref{gain_B}) stem from how the creation and annihilation operators act on many-body Fock states. 

\subsubsection{Loss}
The dissipator $D_{\rm loss}[P]$ describes the spontaneous decay of a boson  from any occupied state to elsewhere: 
\begin{eqnarray}
  D_{\rm loss}[P(\vec{N})]&=&\gamma_{\rm out}\sum_{k,\alpha}
    \left[(N_{k\alpha} + 1)P(\vec N+\vec e_{k\alpha})
    \right. \nonumber \\
&& \left.   \hspace{1.2cm}
 -N_{k\alpha}P(\vec N)\right],
\label{loss_A}
\end{eqnarray}
where the first term on the right-hand side represents a transition from a configuration containing one extra particle in the $(k,\alpha)$ mode. The loss of a particle from this mode then brings the system to the target configuration $\vec{N}$. The second term describes a process starting from the  configuration $\vec{N}$, so that a particle loss from this state decreases the probability $P(\vec{N})$. 

\subsubsection{Thermal relaxation}
Thermal relaxation redistributes bosons among different states according to the jump operators (\ref{thermal_jump}):
\begin{align}
D_R[P(\vec{N})]=&
\sum_{k\neq k'}\sum_{\alpha,\alpha'}
\Gamma_{k\alpha\to k'\alpha'}\,
\Bigl[
 (N_{k\alpha}+1)N_{k'\alpha'}\, \notag \\
& \hspace{-1.5cm}
\times
    P\bigl(\vec N+\vec e_{k\alpha}-\vec e_{k'\alpha'}\bigr) 
-N_{k\alpha}(N_{k'\alpha'}+1)\,P(\vec N)
\Bigr],
\label{TR_dissipator}
\end{align}
where the first term on the right-hand side represents a transition containing $N_{k\alpha}+1$ particles in state $(k,\alpha)$ and $N_{k'\alpha'}-1$ particles in state $(k',\alpha')$. The transfer of a particle from the former state to the latter then brings the system to the target configuration $\vec{N}$. The second term describes a process starting from the configuration $\vec{N}$, so that the same transfer of a boson decreases the probability $P(\vec{N})$.

\subsection{Breaking detailed balance}

In the absence of gain and loss, the thermal-relaxation term $D_{R}[P]$ should bring the system to thermal equilibrium. 
In fact, when $\gamma_{\rm in}=\gamma_{\rm out}=0$, the detailed balance between every pair of transition rates $\Gamma$ leads to the grand-canonical distribution
\begin{equation}
  P^{\rm eq}_{\beta,\mu}(\vec N)
  =\Xi_{\beta,\mu}^{-1}\exp \bigl[\beta(\mu\vec{1}-\vec\epsilon) \cdot \vec N\bigr],
 \end{equation}
which is a stationary solution of Eq.~\eqref{eq:MasterP}, where
\begin{equation}
  \Xi_{\beta,\mu}=\prod_{k,\alpha}
   \frac{1}{1-e^{\beta(\mu-\epsilon_k)}},
\end{equation}
is the grand partition function, and $\vec{1}$ is the all-ones vector. Consequently, the mean occupation number of mode $(k,\alpha)$ is given by the BE distribution:
\begin{equation}
\langle N_{k\alpha}\rangle=
 \frac{1}{e^{\beta(\epsilon_k-\mu)}-1}.
\end{equation}
Thus, our theory naturally incorporates thermal relaxation toward the grand-canonical ensemble which, however, undergoes significant modification in the presence of pump and loss. 

The pump and loss terms drive the system out of thermal equilibrium in
two important ways:  
(i)~$D_{\mathrm{pump}}[P]$ acts as a localized source that feeds particles exclusively into the single level $k_G$, and  
(ii)~$D_{\mathrm{loss}}[P]$ provides a uniform sink that removes particles from all occupied levels.  
Their coexistence breaks detailed balance and establishes a steady
probability flow from the source toward lower-energy levels.  
The Bose-enhancement factors $(N_{k\alpha}+1)$ inherent in both terms
amplify this flow, providing the mechanism for the macroscopic
occupation that we will analyze in the subsequent sections.

\section{Rate equation for energy-level populations}
\label{sec:RateEq}

To analyze the macroscopic occupation, it is sufficient to determine the mean occupation numbers of the single-particle energy levels. To this end, we derive  a closed set of rate equations for these quantities, which will be utilized to analyze the stationary states in Sec.~\ref{sec:OOBE condensation}.  

\subsection{Mean occupation number}

The mean occupation number of mode $(k,\alpha)$ is given by
\begin{align}
 \mN_{k\alpha}(t)=  \sum_{\vec N}N_{k\alpha}\,P(\vec N,t).
\end{align}
The total occupation number of energy level $k$ at time $t$ is given by the sum of $\mathcal{N}_{k\alpha}(t)$ over the degenerate sublevels $\alpha=1,\dots,M_k$:
\begin{align}
 \mN_k(t)= \sum_{\alpha=1}^{M_k}\mN_{k\alpha}(t)
 =\sum_{\vec N}\sum_{\alpha=1}^{M_k}N_{k\alpha}\,P(\vec N,t).
  \label{eq:levelOcc}
\end{align}

Multiplying the master equation~\eqref{eq:MasterP} by $N_{k\alpha}$ and summing over $\vec N$ and $\alpha$, we obtain the rate equation for $\mathcal{N}_k(t)$ (see Appendix~\ref{app:mean-occupation-derivation} for details):
\begin{eqnarray}
  \frac{d\mN_k}{dt}&=&
 \sum_{k'(\neq k)}\sum_{\alpha=1}^{M_k}\sum_{\alpha'=1}^{M_{k'}}
 \left[
 \Gamma_{k'\alpha' \to k\alpha}
    \bigl\langle (N_{k\alpha}+1)N_{k'\alpha'}\bigr\rangle  
    \right. \nonumber\\
 &&\left. 
 \hspace{2cm} -\Gamma_{k\alpha \to k'\alpha'}
    \bigl\langle N_{k\alpha}(N_{k'\alpha'}+1)\bigr\rangle
    \right]
    \nonumber\\
 &&\quad -\gamma_{\rm out}\,\mN_k
 +\delta_{k,k_G}\,M_{k_G}\,\gamma_{\rm in}.
  \label{eq:NumberEqExact}
\end{eqnarray}
The first two terms on the right-hand side arise from the Bose-enhanced
particle fluxes between distinct levels, whereas the remaining two terms
account for particle loss and pump injection. 

\paragraph{Conservation law.}
Summing Eq.~\eqref{eq:NumberEqExact} over $k$, we find that the first two terms on the right-hand side cancel out, yielding
\begin{equation}
 \frac{d}{dt}\sum_{k}\mN_k=
  -\gamma_{\rm out}\sum_{k}\mN_k+\gamma_{\rm in}M_{k_G}.
\end{equation}
Thus, the time derivative of the total particle number on the left-hand side is given by the rate of particle injection minus the rate of particle loss. In a steady state, the left-hand side vanishes, yielding
\begin{align}
  \mN^{\rm ss}_{\rm total} :=\sum_{k}\mN_k^{\rm ss}=M_{k_G}\frac{\gamma_{\rm in}}{\gamma_{\rm out}}.
\label{eq:NtotalSS}
\end{align}
This result implies that the total particle number in the steady state is  tunable via the ratio $\gamma_{\rm in}/\gamma_{\rm out}$.

\subsection{Decoupling approximation}
Equation~\eqref{eq:NumberEqExact} involves second-order correlators. To proceed with the calculation, we assume that the transition rates are
independent of the degenerate-sublevel indices $\alpha,\alpha'$, i.e., 
$\Gamma_{k\alpha\to k'\alpha'}=\Gamma_{k\to k'}$.  For a dilute gas with rapidly relaxing correlations, we can safely neglect the inter-energy-level correlations and factorize the correlation functions as
\begin{equation}
\langle N_{k\alpha}N_{k'\alpha'}\rangle\approx
  \langle N_{k\alpha}\rangle\langle N_{k'\alpha'}\rangle
  =\frac{\mN_k}{M_k}\frac{\mN_{k'}}{M_{k'}}.
\end{equation}
Under this approximation , the rate equation for each level $k$ closes in the mean-field form:
\begin{eqnarray}
  \frac{d\mN_k}{dt}
  &=&\sum_{k'(\neq k)}\left[\Gamma_{k'\to k}
       (\mN_k+M_k)\mN_{k'}    \right.\notag\\
  & &\left. \hspace{0.7cm} -\Gamma_{k\to k'}
       \mN_k(\mN_{k'}+M_{k'}) \right] \notag\\
  & &\quad-\gamma_{\rm out}\,\mN_k
        +\delta_{k,k_G}\,M_{k_G}\,\gamma_{\rm in}.
  \label{eq:RateEq}
\end{eqnarray}
Equation~\eqref{eq:RateEq} contains only the total level populations $\mN_{k}$ and their degeneracies $M_k$, where the latter preserves the characteristics of Bose stimulation within the decoupling approximation.

In the absence of gain and loss (i.e., $\gamma_{\rm out}=\gamma_{\rm in}=0$), Eq.~(\ref{eq:RateEq}) must reduce to the equilibrium BE distribution $\mN_k^{\rm eq}$. Indeed, by imposing the KMS condition~\eqref{eq:KMS} on the transition rates, we obtain
\begin{equation}
e^{-\beta\epsilon_k}\left(1+\frac{M_k}{\mN_k}\right)
=
e^{-\beta\epsilon_{k'}}\left(1+\frac{M_{k'}}{\mN_{k'}}\right)
\end{equation}
Since this equation holds identically for all $k$ and $k'$,it must be constant  with respect to both indices. Equating this constant to $e^{-\beta\mu}$, with $\mu$ being a parameter to be identified below,  we arrive at
\begin{equation}
\mN_k^{\rm eq}=\frac{M_k}{e^{\beta(\epsilon_k-\mu)}-1}.
\label{BE_distribution}
\end{equation}
Thus, we obtain the BE distribution as the stationary solution of Eq.~\eqref{eq:RateEq}: $d\mN_k/dt=0$. We also confirm that the parameter $\mu$ introduced above can be interpreted as the chemical potential in the absence of gain and loss.

In the next section, we analyze the stationary solutions
of Eq.~\eqref{eq:RateEq} in the presence of pump and loss, which enables
a quantitative description of genuinely out-of-equilibrium BE condensation in a driven-dissipative harmonic trap.

\section{Out-of-equilibrium Bose-Einstein Condensation}
\label{sec:OOBE condensation}

In what follows, we focus on an isotropic harmonic potential as it reproduces, to an excellent approximation, the experimental setup realized by Morita \textit{et al.} for paraexcitons in Cu$_2$O,  where 
a strain–induced harmonic confinement was employed to observe an exciton BE condensate~\cite{Morita2022}.  

The single-particle states are labeled by $(k,\alpha)$, where $k=0,1,\dots$ denotes the harmonic-oscillator levels and $\alpha$ runs over the $M_k$ degenerate sublevels.  In $d$ spatial dimensions, the degeneracy is given by
\begin{align}
  \epsilon_{k,\alpha} &= \hbar\omega\!\left(k+\frac{d}{2}\right),
  \label{eq:HOlevels}\\
  M_k &= \binom{k+d-1}{d-1}.
  \label{eq:Degeneracy}
\end{align}
To satisfy the KMS condition \eqref{eq:KMS}, the bath-induced transition rates are chosen as
\begin{equation}
  \Gamma_{k+1\to k}= \gammath,\qquad
  \Gamma_{k\to k+1}= \gammath\,e^{-\beta\hbar\omega}.
  \label{eq:RatesHO}
\end{equation}
Substituting Eqs.~\eqref{eq:HOlevels}–\eqref{eq:RatesHO} into Eq.~\eqref{eq:RateEq} and assuming that the transitions occur only to nearest-neighbor energy levels, we obtain the following equation for the stationary state:
\begin{align}
  0 &= e^{-\beta\hbar\omega}\!
      \left[-\mN_{k}(\mN_{k+1}+M_{k+1})+(\mN_{k}+M_k)\mN_{k-1}\right]
      \notag\\
    &\quad+(\mN_{k}+M_k)\mN_{k+1}
                 -\mN_{k}(\mN_{k-1}+M_{k-1}) \notag\\
    &\quad-\gout\,\mN_{k}
      +\delta_{k,k_G}\,M_{k_G}\,\gin,
  \label{eq:SSrateHO}
\end{align}
where
\begin{equation}
\gout :=\frac{\gammaout}{\gammath}, \ \ \
\gin :=\frac{\gammain}{\gammath}
\label{gout_gin}
\end{equation}
are the dimensionless parameters.
We note that the pumping is restricted to the highest-energy level
$k=k_G\gg1$ with rate $\gamma_{\rm in}$, whereas loss occurs at every energy level at a constant rate $\gamma_{\rm out}$. The ratios of these rates against the thermalization rate $\gamma_{\rm R}$ constitute the key parameters for studying the out-of-equilibrium BE condensation.

\subsection{Continuum approximation}

In the experiment of Ref.~\cite{Morita2022}, highly excited excitons are created by a laser field, implying that the pump injects particles far above the ground state, i.e., $k_G\gg1$.  We use this fact to normalize the level index $k$ by $k_G$:
\begin{equation}
  x:=\frac{k}{k_G}\in[0,1], \ \ \  dx=\frac{1}{k_G}.
  \label{eq:ContLimit1}
\end{equation}
Then, away from the boundaries, the level population $\mathcal{N}_k$ can be treated as a smooth function of the quasi-continuous variable $x$:
\begin{equation}
  n(x):=\mN_k.
  \label{eq:ContLimit2}
\end{equation}
Crucially, the ground-state population $\mN_0$ must be singled out and treated separately, as the lower-edge value of the smooth continuum distribution does not necessarily coincide with $\mN_0$.
We also have 
\begin{equation}
\mN_{k\pm1}=n(x)\pm n'(x)\,dx+{\cal O}(dx^{2}), 
\end{equation}
and
\begin{equation}
M(x):=M_k=\binom{k_Gx+d-1}{d-1}.
\label{M(x)}
\end{equation}
Substituting Eqs. (\ref{eq:ContLimit2})-(\ref{M(x)}) into Eq.~\eqref{eq:SSrateHO} and expanding the terms up to leading orders, we obtain
\begin{align}
&\frac{d}{dx}\Bigl[n(x)\bigl(n(x)+M(x)\bigr)\Bigr]\notag\\
&\quad- k_G \goutb\,n(x)
+\ginb\,M(1)\,\delta(x-1)=0,
\label{eq:ContinuumEq}
\end{align}
where $k_G\delta_{k,k_G}\rightarrow\delta(x-1)$, and we have introduced the temperature-normalized dimensionless rates for pump and loss:
\begin{equation}
\ginb  := \frac{\gin }{1-e^{-\beta\hbar\omega}}, \qquad
  \goutb := \frac{\gout}{1-e^{-\beta\hbar\omega}}.
  \label{rescaling}
\end{equation}
Thus, Eq.~\eqref{eq:ContinuumEq} is expressed solely in terms of
$\goutb$, $\ginb$, and $k_G$. 
The first term in Eq.~(\ref{eq:ContinuumEq}) describes the Bose-enhanced
net particle transfer along the harmonic ladder, the second accounts for
particle loss, and the third describes the injection of particles at the source level $x=1$. 
Physically, the rescaling in Eq.~(\ref{rescaling}) quantifies the gain and loss relative to the difference between the downward and upward transition-rate coefficients, \(\gamma_R(1-e^{-\beta\hbar\omega})\), which governs the proportion of the net downward particle current in the leading-order continuum approximation.
All subsequent results—including the density profiles $n(x)$, the non-condensed population $\mN_{\rm nc}$, and condensate fraction $f_0$—will be expressed in terms of  these reduced variables.  

\subsection{Identification of non-equilibrium BE condensation}

We first define the number of bosons in the excited states $\mN_{\rm nc}$, which merges into a quasi-continuum in the continuum limit:
\begin{equation}
\mN_{\rm nc}    = \sum_{k=1}^{\infty} \mN_k.
 \label{eq:ThermalDensity0}
\end{equation}
Assuming that boundary contributions and the population above the pumped shell are subleading, we obtain the following continuum approximation:
\begin{equation}
\mN_{\rm nc} \simeq k_G
\int_{0}^{1}\!dx\,n(x). 
\label{eq:ThermalDensity}
\end{equation}
If low-lying excited levels carry a leading-order population, they must be retained in $\mN_{\rm nc}$, as discussed below for two and three dimensions (see Eqs. (64)–(66) and the discussion following Eq. (71)).  

The noncondensate fraction $f_{\mathrm{nc}}$ is defined as the ratio of this number to the total number of bosons given in Eq.~\eqref{eq:NtotalSS}:
\begin{align}
    f_{\mathrm{nc}} := \frac{\mN_{\rm nc}}{\mN^{\rm ss}_{\rm total}}.
\end{align}
The condensate fraction is defined by
\begin{align}
    f_0 := 1- f_{\rm nc}.
\end{align}
A nonzero value of $f_0$ as the total population becomes macroscopically large implies that a macroscopic number of bosons occupy the ground state, which fulfills the condition for BE condensation.

At thermal equilibrium, the onset of BE condensation is determined solely by the chemical potential $\mu$ reaching the lowest
single-particle energy.
Far from equilibrium,  by contrast, the criterion $f_0>0$ is governed
by the dynamic balance between particle injection at the pump level and uniform loss across the entire spectrum. Without recourse to a global chemical potential, and with temperature alone being insufficient to predict the condensate fraction, the pump–to–loss ratio emerges as the central nonequilibrium control knob. This ratio renders the condensate fraction tunable in real time, fundamentally distinguishing the system from its equilibrium counterpart.

Defining the number of condensate bosons as 
\[
\mN_0=\mN^{\rm ss}_{\rm total}-\mN_{\rm nc},
\]
the condensate fraction is given from Eq.~\eqref{eq:NtotalSS} as
\begin{equation}
  f_0=\frac{\mN_0}{\mN^{\rm ss}_{\rm total}}
  =1-\frac{\gout\,\mN_{\rm nc}}{M_{k_G}\gin}.
  \label{eq:CondFraction}
\end{equation}
Here $\mN_{\rm nc}$ must be obtained from the stationary distribution; no pump-independent saturation value or finite critical pump is assumed.
In the following sections we evaluate this fraction in each dimension and compare the three-dimensional results
with the experimental ones reported in Ref.~\cite{Morita2022}.

\section{Condensate fraction}
\label{sec:CF}

We now solve Eq.~\eqref{eq:ContinuumEq} to find the steady–state condensate fraction $f_0$ in one, two and three spatial dimensions.

\subsection{One dimension}
\label{subsec:1D}

In one dimension, every harmonic level is non-degenerate (i.e., $M(x)=1$), which reduces Eq.~\eqref{eq:ContinuumEq} to a single, first-order ordinary differential equation.  
Its unique steady-state solution can be written in a closed form using the
Lambert $W$ function,
\begin{equation}
  n(x)=\frac12\,W\!\Bigl[2n_1\,  e^{\,2n_1- k_G\goutb(1-x)}\Bigr],
         \ \ \  0\le x<1,
  \label{eq:1Dprofile_final}
\end{equation}
where the population density just below the pumped shell, $n_1:= n(1^-)$, is determined to the leading order from Eq.~(\ref{eq:SSrateHO}) as
\begin{equation}
  \ginb = n_1\bigl(n_1+1\bigr).
  \label{eq:1Dbc_final}
\end{equation}
Equation~\eqref{eq:1Dprofile_final} reflects the nonlinearity introduced by bosonic stimulation; the gradient of the occupation profile depends nonlinearly on the local occupation, as shown below in Eq. (\ref{gradient}).

We use the obtained solution~\eqref{eq:1Dprofile_final} to investigate the properties of the condensate in the limit $k_G\to\infty$ with $k_G\goutb$, $\ginb$, and $q:=e^{-\beta\hbar\omega}<1$ held fixed. In this regime, the discrete ground state can attain an $O(k_G)$ occupation, while corrections from the boundary regions and the population above the pump remain subleading. It follows from Eq~(\ref{eq:ContinuumEq}) with $M(x)=1$ that the particle density for $0<x<1$ obeys
\begin{equation}
\frac{dn}{dx}= k_G\goutb\frac{n}{1+2n}.
\label{gradient}
\end{equation}
This equation can be used to change the integration variable (\ref{eq:ThermalDensity}) from $x$ to $n$. Then the resulting integration gives the noncondensed population as
\begin{eqnarray}
  \mN_{\mathrm{nc}}^{(1\mathrm{D})}
     & \simeq&\frac{1}{\goutb}\Bigl[n_1(n_1+1)-n_{\mathrm e}(n_{\mathrm e}+1)\Bigr] \nonumber \\
     &=&\frac{1}{\goutb}\Bigl[\ginb-n_{\mathrm e}(n_{\mathrm e}+1)\Bigr],
  \label{eq:nth1D_correct}
\end{eqnarray}
where
$n_{\mathrm e}:=\lim_{x\to0^+}n(x)$ is the lower-edge value of the smooth excited-state distribution, with
$  n_{\mathrm e}
         =\tfrac12 W\!\bigl[2n_1\,e^{2n_1- k_G\goutb}\bigr]$ and we use Eq.~(\ref{eq:1Dbc_final}) in obtaining the second equality.  Since the total number of bosons in  one dimension is given from Eq.~(\ref{eq:NtotalSS}) with $M_{k_G}=1$ by
\begin{equation}
\mN^{(1D)}=\frac{\ginb}{\goutb},
\label{totalN_1D}
\end{equation}
the number of condensate bosons is given by
\begin{eqnarray}
  \mN_0^{(1\mathrm{D})}= \mN^{(1D)}- \mN_{\mathrm{nc}}^{(1\mathrm{D})}
     \simeq\frac{n_{\mathrm e}(n_{\mathrm e}+1)}{\goutb}.
  \label{1D_N_0}
\end{eqnarray}
It is noteworthy that the number of condensate bosons is not equal to  $n_{\mathrm e}$ but depends nonlinearly on it. Here, $n_{\mathrm e}$ represents an extrapolated continuum value rather than the occupation of the discrete ground state or the first excited level. As noted above, $ \mN_0^{(1\mathrm{D})}$ is determined by subtracting the sum over the excited states from the total number of bosons. 
This distinction is consistent with the discrete rate equation. In units of $\gamma_R$, the downward net current between adjacent levels is
\begin{equation}
J_k=(\mN_{k-1}+1)\mN_k-q\mN_{k-1}(\mN_k+1),
\end{equation}
and the ground-state balance yields $J_1=\gout\mN_0^{(1\mathrm D)}$,  since the incoming current from the first excited level
must balance particle loss from the ground state, which is not directly pumped (see also Eq. ~(\ref{eq:RateEq})).

 Within the low-energy boundary region, losses from each excited level are subleading, which implies $J_k\simeq J_1$ and
\begin{equation}
\mN_k\simeq\frac{J_1+q\mN_{k-1}}{1+(1-q)\mN_{k-1}}.
\end{equation}
The stable positive fixed point of this recurrence is $n_{\mathrm e}$, corresponding to $J_1\simeq(1-q)n_{\mathrm e}(n_{\mathrm e}+1)$. Matching this solution to the smooth continuum therefore yields Eq.~\eqref{1D_N_0}.

The condensate fraction $f_{0}^{(1\mathrm{D})}$ is defined as the ratio of $ \mN_0^{(1\mathrm{D})}$ to $\mN^{(1D)}$:
\begin{equation}
  f_{0}^{(1\mathrm{D})}=\frac{\mN_0^{(1\mathrm{D})}}{\mN^{(1D)}}
     \simeq\frac{n_{\mathrm e}(n_{\mathrm e}+1)}{\ginb}.
  \label{eq:f01D_correct}
\end{equation}
Thus, the condensate fraction is determined to leading order by $\ginb$ and $n_{\mathrm e}$, where $n_{\mathrm e}$ is given by Eq.~(\ref{eq:1Dprofile_final}).

The lower-edge value $n_{\mathrm e}$ is suppressed by leakage.  In fact, since $W(z)\sim z$ for $|z|\ll1$ and $n_1\simeq \ginb$ from Eq.~(\ref{eq:1Dbc_final}) for $\ginb\ll1$, we obtain
\begin{equation}
  n_{\mathrm e}\simeq n_1\,e^{\,2n_1- k_G\goutb}
          \simeq \ginb\,e^{\,2\ginb- k_G\goutb}.
\label{n_0}
\end{equation}
Defining $G:=\ginb$ and $K:=k_G\goutb$, the leading-order condensate fraction approaches $e^{-K}$ as $G\to0^+$ in the continuum limit. While this fraction is exponentially small for $K\gg1$, no finite critical pump exists. For $G\gg\max(1,K^2)$, the same solution gives
\begin{equation}
\mN_{\rm nc}^{(1\mathrm D)}\sim k_G\sqrt{G},
\qquad 1-f_0^{(1\mathrm D)}\sim\frac{K}{\sqrt{G}}.
\end{equation}
Thus, the excited-state population continues to grow, albeit more slowly than the total population $\mN^{(1D)}=k_GG/K$, causing the condensate fraction  to approach unity via a continuous crossover rather than excited-state  saturation.
This mechanism is unique to driven–dissipative systems, in which a nonzero condensate fraction arises from the interplay of pumping, loss, and bosonic stimulation without undergoing a true phase transition.

\subsection{Two dimensions}
\label{subsec:2D}

For a two–dimensional isotropic harmonic trap, the energy-shell degeneracy (\ref{M(x)}) is given by
\begin{equation}
M(x)=k_{G}x+1.
\label{2D_M(x)}
\end{equation}
To eliminate the dependence on the remote pump shell $k_{G}$, we introduce the shell–normalized variables
\begin{equation}
  \bar n(x):=\frac{n(x)}{k_{G}},\qquad
  \ginbbar:=\frac{\ginb}{k_{G}},\qquad
  \goutbbar:=\goutb ,\label{eq:rescaled_2D}
\end{equation}
so that $\ginbbar$ represents the temperature-renormalised pump rate per shell, while $\goutbbar$ denotes the temperature-renormalized loss.  Neglecting terms of the order of $k_G^{-1}$, we obtain the continuum steady-state equation (\ref{eq:ContinuumEq}) as
\begin{equation}
  \frac{d}{dx}\!\Bigl[\bar n\bigl(\bar n+x\bigr)\Bigr]
  -\goutbbar\,\bar n
  +\ginbbar\,\delta(x-1)=0 .
  \label{eq:2Dfull}
\end{equation}

Everywhere except at the pumped shell ($0\le x<1$), the $\delta$-function source vanishes.  Retaining the nonlinear current yields
\begin{equation}
\frac{d\bar n}{dx}=\frac{(\goutbbar-1)\bar n}{2\bar n+x}.
\label{eq:2Dode}
\end{equation}
The quadratic term can be neglected only in the dilute part of the thermal wing where $\bar n\ll x$, reducing the equation to
\begin{equation}
x\frac{d{\bar n}}{dx}=(\goutbbar-1)\bar n.
\end{equation}
Under this local approximation, the solution exhibits an algebraic tail,
\begin{equation}
  \bar n(x)\simeq A\,x^{-(1-\goutbbar)},\qquad \bar n(x)\ll x,
  \label{2D_solution}
\end{equation}
where the prefactor $A$ must be determined by matching to the full solution. For $0<\goutbbar<1$, however, this approximation breaks down  near $x=0$, and the resulting profile cannot be integrated down to the lower edge to obtain the noncondensed population. We therefore employ Eq.~\eqref{eq:2Dode} to describe the smooth continuum profile.

Integrating the steady-state equation over an infinitesimal interval
around $x=1$ determines $\bar n_1\equiv\bar n(1^-)$ via
\begin{equation}
  \bar n_{1}\bigl(\bar n_{1}+1\bigr)=\ginbbar.
\end{equation}
The solution is
\begin{equation}
  \bar n_{1}=\frac{\sqrt{1+4\ginbbar}-1}{2},
\end{equation}
which scales linearly with the reduced pump for 
$\ginbbar\ll1$ and crosses over to $\bar n_{1}\simeq\sqrt{\ginbbar}$ for 
$\ginbbar\gg1$.

The smooth continuum contributes $k_G^2\int_0^1\bar n(x)\,dx$ to the total particle number, so that in two dimensions it is given by Eq.~\eqref{eq:NtotalSS} with $M_{k_G}=k_G+1$:
\begin{equation}
\mN^{(2D)}=(k_G+1)\frac{\ginb}{\goutb}
\simeq k_G^2\frac{\ginbbar}{\goutbbar}.
\label{totalN_2D}
\end{equation}
Integrating the full continuum equation yields the residual weight
\begin{equation}
R_2:=1-\frac{\goutbbar}{\ginbbar}\int_0^1\bar n(x)\,dx
=\frac{\bar n(0^+)^2}{\ginbbar}.
\label{eq:R2}
\end{equation}
In general, this quantity does not equal the ground-state fraction, as the few lowest-lying excited states contributing to the population are excluded from the smooth continuum description. 
For example, at $\ginbbar=\goutbbar=0.1$, the full continuum description yields $R_2=0.83960$, whereas formally extending the dilute-tail solution~\eqref{2D_solution} with $A=\bar n_1$ down to $x=0$ gives only $0.08392$. This extrapolation violates the condition $\bar n\ll x$ near the lower edge and cannot be used to determine the condensate fraction.

To distinguish the ground state from low-lying excited levels, we solve the discrete equation~\eqref{eq:SSrateHO} with the exact degeneracies~\eqref{eq:Degeneracy} for $k=0,\ldots,k_{\max}$, where $k_{\max}>k_G$. In units of $\gamma_R$, the downward current is given by
\begin{equation}
J_k=(\mN_{k-1}+M_{k-1})\mN_k
-q\mN_{k-1}(\mN_k+M_k),
\label{eq:discrete-current}
\end{equation}
where $q=e^{-\beta\hbar\omega}$. The stationary equations read
\begin{equation}
J_{k+1}-J_k-\gout\mN_k
+\delta_{k,k_G}M_{k_G}\gin=0,
\label{eq:discrete-balance}
\end{equation}
subject to the boundary conditions $J_0=J_{k_{\max}+1}=0$. We increase the upper cutoff until the ground-state fraction converges. For each numerical solution, we require the residual $r_k$, defined as the left-hand side of Eq.~\eqref{eq:discrete-balance}, to satisfy $\frac{\max_{0\le k\le k_{\max}}|r_k|}{M_{k_G}\gin}<2\times10^{-12}$, and verify global particle-number balance via $\left|\frac{\gout\sum_{k=0}^{k_{\max}}\mN_k}{M_{k_G}\gin}-1\right|<10^{-9}$, where the sum includes all levels from the ground state to the upper cutoff $k_{\max}>k_G$. The condensate fraction is evaluated directly as
\begin{equation}
f_0^{(d\mathrm D)}=\frac{\mN_0}{\sum_{k=0}^{k_{\max}}\mN_k}
=\frac{\gout\mN_0}{M_{k_G}\gin}.
\label{eq:discrete-f0}
\end{equation}
This procedure is also used in three dimensions below.

The need to treat the lowest-energy levels separately is already apparent from the first few states. For $\mN_0\gg\mN_1,1$, setting $k=0$ in Eq.~(\ref{eq:SSrateHO}) yields
\begin{equation}
\mN_1\simeq\frac{\gout+qd}{1-q},
\end{equation} 
and subsequently setting $k=1$ gives
\begin{equation}
\frac{\mN_2}{\mN_0}\simeq\frac{\gout}{\gout+d(1+q)}.
\label{eq:boundary-ratio}
\end{equation}
In 1D, $\gout\propto k_G^{-1}$, causing this ratio to vanish. In 2D, by contrast, $\gout=(1-q)\goutbbar$ remains fixed, and thus the low-lying excited levels are not necessarily subleading. Consequently, the discrete ground-state fraction depends on $q$ and $k_G$ in addition to the reduced rates; the continuum residual alone does not establish a universal condensate fraction. Numerical examples are provided in Table~\ref{tab:discrete-check} below.

\subsection{Three dimensions}
\label{subsec:3D}

In three dimensions, the shell degeneracy grows quadratically as 
$M(x)\simeq\tfrac12 k_{G}^{2}x^{2}$ away from the low-energy boundary.  To work with dimensionless quantities, we rescale each extensive variable by the appropriate power of $k_{G}$, introducing
\begin{equation}
  \bar{n}(x)=\frac{n(x)}{k_{G}^{2}},\qquad
  \ginbbar=\frac{\ginb}{k_{G}^{2}},\qquad
  \goutbbar=\frac{\goutb}{k_{G}} .
  \label{eq:rescaled_3D}
\end{equation}
These scalings are introduced in the same spirit as in Eq.~\eqref{eq:rescaled_2D}, differing only in spatial dimensionality.
In these units, the continuum steady–state equation
\eqref{eq:ContinuumEq} becomes
\begin{equation}
  \frac{d}{dx}\!\Bigl[\bar{n}\bigl(\bar{n}+\tfrac12 x^{2}\bigr)\Bigr]
  -\goutbbar\,\bar{n}
  +\tfrac12\ginbbar\,\delta(x-1)=0 ,
  \label{eq:3Dfull}
\end{equation}
with $x\in[0,1]$.  
As in two dimensions, the term $\bar{n}^{2}$ cannot be neglected at small~$x$, where it competes with the rapidly growing geometric factor $x^{2}$.  We therefore solve Eq.~\eqref{eq:3Dfull} numerically for the smooth continuum profile and use the discrete equations~\eqref{eq:discrete-balance} and \eqref{eq:discrete-f0} to determine the ground-state fraction.

\begin{figure}
    \centering
    \includegraphics[width=\columnwidth]{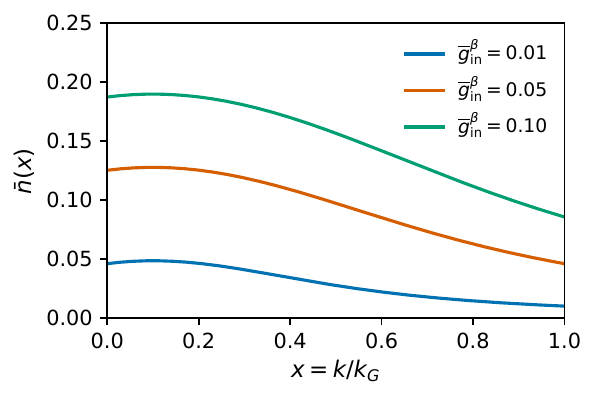}
    \includegraphics[width=\columnwidth]{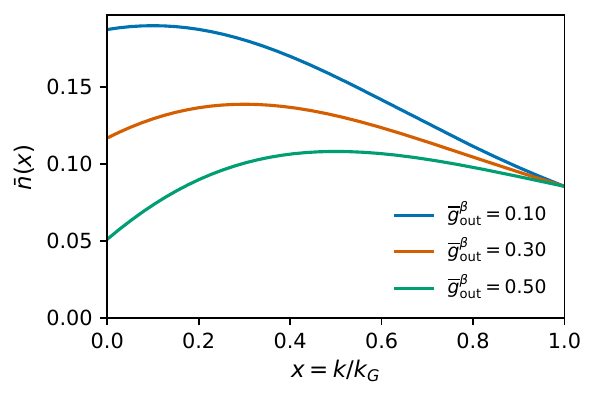}
    \caption{Steady-state profiles of the shell–normalised occupation 
$\bar{n}(x)=n(x)/k_G^{2}$ in $d=3$, plotted against \ $x=k/k_G$, where
$x=0$ denotes the lower continuum edge and $x=1$ the pumped shell.
The upper panel shows results for a fixed loss $\goutbbar=0.10$ and three values of the pump rate: $\ginbbar$=0.01 (blue), 0.05 (orange), and 0.10 (green). Increasing $\ginbbar$ lifts the entire curve upward and enhances the low-$x$ weight.
The lower panel shows results for a fixed pump rate $\ginbbar=0.10$ with $\goutbbar$=0.10 (blue), 0.30 (orange), and 0.50 (green). Here increasing loss lowers the profile, reducing its low-energy weight. Note that these curves represent continuum profiles rather than the discrete ground-state occupations.}
    \label{fig:profiles-3D}
\end{figure}

Solving Eq.~\eqref{eq:3Dfull} requires specifying the dimensionless occupation $\bar{n}_{1}\equiv\bar{n}(1^{-})$ just below the pumped shell.
Particle-current balance across the infinitesimal layer that hosts the
$\delta$–function source determines this value through
\begin{equation}
\bar{n}_{1}\bigl(\bar{n}_{1}+\frac{1}{2}\bigr)=\frac{1}{2}\ginbbar.
\end{equation}
We integrate the first–order ordinary differential equation toward $x=0$ using an adaptive Runge–Kutta method. The resulting steady–state profiles (Fig.~\ref{fig:profiles-3D}) are nonmonotonic for $0<\goutbbar<1$ and exhibit a maximum at $x=\goutbbar$, which follows directly from the differential equation:
\begin{equation}
\frac{d\bar{n}}{dx}=\frac{2(\goutbbar-x)\bar{n}}{4\bar{n}+x^2}. 
\end{equation}
The maximum reflects the competition between particle loss and the energy dependence of the degeneracy. 

The upper panel of Fig.~\ref{fig:profiles-3D} shows the result for a fixed loss
$\goutbbar=0.10$ and three values of $\ginbbar$.  Increasing the pump rate shifts the entire curve upward and enhances the low-energy population.  The lower panel shows the case of a fixed pump rate $\ginbbar=0.10$ with varying loss.  Here, increasing $\goutbbar$ shifts the peak toward higher $x$, illustrating how leakage hampers the buildup of low–energy occupation.  Integrating Eq.~\eqref{eq:3Dfull} yields the continuum residual
\begin{equation}
R_3:=1-\frac{2\goutbbar}{\ginbbar}\int_0^1\bar n(x)\,dx
=\frac{2\bar n(0^+)^2}{\ginbbar}.
\label{eq:R3}
\end{equation}
As with $R_2$, a nonzero $R_3$ alone does not establish the presence of a ground-state condensate. At fixed $\goutbbar$, the discrete loss rate scales as $\gout=(1-q)k_G\goutbbar$. Consequently, the ratio in Eq.~\eqref{eq:boundary-ratio} approaches unity with increasing $k_G$,  provided that the large-ground-occupation approximation holds. Thus, low-lying excited levels must be treated separately rather than lumped into the ground state.

\begin{figure}
    \centering
    \includegraphics[width=\columnwidth]{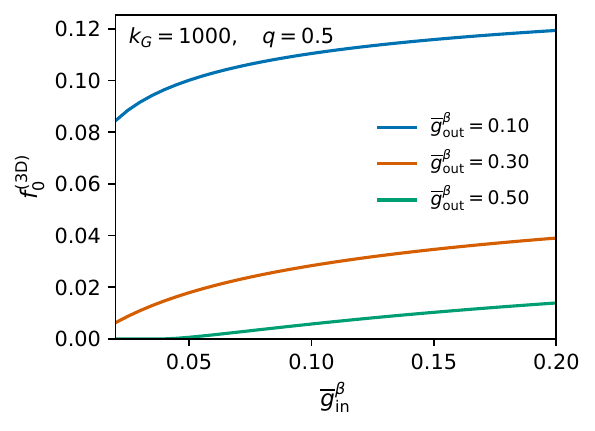}
    \caption{Condensate fraction $f^{(3\mathrm{D})}_{0}$ in three dimensions as a function of the temperature–renormalized pump rate
$\ginbbar$ for three fixed loss rates $\goutbbar$=0.10 (blue), 0.30 (orange), 0.50 (green). The ground occupation is computed from the discrete rate equation at $k_G=1000$ and $q=e^{-\beta\hbar\omega}=0.5$ using exact degeneracies. The upper cutoff is set to $k_{\max}=k_G+200$; increasing it to $k_G+400$ alters the condensate fraction by less than $10^{-8}$ at both endpoints of each curve. These parameters are not fitted to the exciton experiment.
}
\label{fig:f0-3D}
\end{figure}

Figure~\ref{fig:f0-3D} shows the condensate fraction $f_{0}^{(3\mathrm{D})}$ as a function of $\ginbbar$, obtained directly from the discrete ground-state occupation using Eq.~\eqref{eq:discrete-f0}.
For a fixed loss rate, $f_{0}^{(3\mathrm{D})}$ rises smoothly with the pump rate, exhibiting no sharp threshold across the finite-system parameter range shown. While increasing loss suppresses the ground-state fraction,  this dependence cannot be inferred from the continuum residual or a single loss-to-gain ratio alone.

Table~\ref{tab:discrete-check} compares the discrete ground fractions with the continuum residuals. In three dimensions, the ground-state fraction decreases substantially with increasing $k_G$ at fixed reduced rates, even though $R_3$ remains unchanged. Thus, these finite-system results alone do not establish a nonzero limiting ground fraction. While the nonlinear current is present in all three dimensions,  what differs is the degeneracy and the contribution of the discrete low-energy region.

\begin{table}
\centering
\begin{tabular}{cccc}
\hline
$d$ & $k_G$ & $f_0$ (discrete) & $R_d$ (continuum) \\
\hline
2 & 100 & 0.82970 & 0.83960 \\
2 & 1000 & 0.81910 & 0.83960 \\
2 & 10000 & 0.81471 & 0.83960 \\
3 & 100 & 0.35733 & 0.70044 \\
3 & 1000 & 0.11043 & 0.70044 \\
3 & 10000 & 0.03287 & 0.70044 \\
\hline
\end{tabular}
\caption{Discrete ground-state fractions and smooth-continuum residuals at $q=0.5$ and $\ginbbar=\goutbbar=0.1$, evaluated using the dimension-dependent rate definitions. Increasing the upper cutoff from $k_G+200$ to $k_G+400$ changes every displayed ground-state fraction by less than $10^{-8}$.}
\label{tab:discrete-check}
\end{table}

\section{Exciton Bose--Einstein condensation}
\label{sec:Cu2O}

The para–exciton condensate observed by Morita \textit{et~al.}\
\cite{Morita2022} provides an ideal testbed for our driven–dissipative theory of BE condensation.
Below, we translate the experimental parameters into the dimensionless
rates entering our three–dimensional model. We then evaluate the discrete ground-state fraction in~\eqref{eq:discrete-f0} and assess its order-of-magnitude consistency with the observed percent-level condensate fraction. This comparison relies on effective rates and an assumed injection energy, and thus does not constitute a parameter-free prediction. The parameters used in  Fig.~\ref{fig:f0-3D} are distinct from those employed below.

\begin{table*}[t]
\centering
\begin{tabular}{|l|c|c|}
\hline
Quantity & $P_{\rm pump}=1.6\;\mathrm{mW}$ & $P_{\rm pump}=8.8\;\mathrm{mW}$ \\
\hline
Generation rate $P_{\rm gen}$ & $2.44\times10^{15}\;\mathrm{s}^{-1}$ & $1.34\times10^{16}\;\mathrm{s}^{-1}$ \\
\hline
Total population $N_{\rm total}$ (approximate) & $1.72\times10^8$ & $4.86\times10^8$ \\
\hline
Effective lifetime $N_{\rm total}/P_{\rm gen}$ & $70.5\;\mathrm{ns}$ & $36.2\;\mathrm{ns}$ \\
\hline
Reduced loss $\goutbbar$ & $0.224$ & $0.437$ \\
\hline
Reduced pump $\ginbbar$ & $0.0139$ & $0.0764$ \\
\hline
Calculated fraction $f_0$ & $0.0208$ & $0.0137$ \\
\hline
Reported maximum fraction (approximate) & $0.013$ & $0.016$ \\
\hline
\end{tabular}
\caption{Phenomenological comparison with Ref.~\cite{Morita2022}. Generation rates follow Eq.~(27) in its Supplementary information; total populations are estimated by summing its rounded thermal and condensate populations. Effective lifetimes are inferred from particle-number balance rather than independently measured decay times. Calculated fractions are evaluated at $T=64\;\mathrm{mK}$,  $k_G=1770$, and a common representative coefficient $\gammath=5\times10^6\;\mathrm{s}^{-1}$. The reported maxima indicate the observed scale rather than a point-by-point fit to the temperature dependence.}
\label{tab:correspondence}
\end{table*}

\subsection{Experimental situations}

Morita et al.~\cite{Morita2022} used a strain–induced harmonic potential with trap frequencies of 8.4, 8.4, and 12.4 MHz along the three principal axes to confine paraexcitons. For our isotropic model, we adopt their geometric mean, $\omega/2\pi\simeq9.56\;\mathrm{MHz}$, which corresponds to $\hbar\omega/k_B\simeq0.459\;\mathrm{mK}$. Setting the bath temperature to be the lowest mixing-chamber temperature, $T=64\;\mathrm{mK}$, yields
\begin{equation}
q=e^{-\hbar\omega/(k_BT)}\simeq0.99285.
\end{equation}
The bath temperature entering detailed balance is not identified with the effective temperature inferred from exciton luminescence, which can differ from the mixing-chamber temperature.
In the experiment, a continuous-wave fibre Raman laser was used to generate dipole-allowed orthoexcitons, which were subsequently converted into paraexcitons.
We provisionally take the injection energy to be $\Delta E=0.07\;\mathrm{meV}$ above the ground state, giving $k_G\simeq\Delta E/\hbar\omega\simeq1770$ and $M_{k_G}=(k_G+1)(k_G+2)/2=1{,}569{,}106$. This injection energy is a model assumption rather than an independently measured value reported in Ref.~\cite{Morita2022}.

Morita \textit{et al.}\ reported thermal populations of approximately $1.7\times10^8$ and $4.8\times10^8$, and condensate populations of approximately $2\times10^6$ and $6\times10^6$ at 1.6 and 8.8 mW, respectively. We use the resulting total populations in Table~\ref{tab:correspondence} solely to set the overall population scale rather than to reproduce their reported maximum condensate fractions of about 0.013 and 0.016.

The 600$\pm$20 ns lifetime reported in Ref.~\cite{Morita2022} characterizes density-independent loss, whereas the 7--12 ns timescale quoted therein corresponds to ortho-to-paraexciton conversion rather than the thermal relaxation coefficient $\gammath$ in our model. At the densities relevant to the condensation experiment, two-body loss also becomes significant. Following Eq.~(27) in Supplementary information of Ref.~\cite{Morita2022}, we estimate the total generation rate from the measured 50\% absorption, assuming complete conversion and trapping:
\begin{equation}
P_{\rm gen}=\frac{0.5P_{\rm pump}}{hc/\lambda_{\rm laser}} \ \  {\rm at} \ \  \lambda_{\rm laser}=606.1\;\mathrm{nm}.
\label{eq:exp-generation}
\end{equation}
To model the net loss with a uniform rate, we set
\begin{equation}
\gamma_{\rm out}^{\rm eff}=\frac{P_{\rm gen}}{N_{\rm total}},\qquad
\gammain=\frac{P_{\rm gen}}{M_{k_G}}.
\label{eq:exp-effective-loss}
\end{equation}
In this comparison, $\gammaout$ is therefore interpreted as an effective loss rate $\gamma_{\rm out}^{\rm eff}$, corresponding to  effective lifetimes of approximately 70.5 and 36.2 ns, respectively. While this replacement preserves the total particle-number balance, it does not resolve the state-dependent effects of two-body loss and renders  the effective loss rate dependent on the pump power. Equation~\eqref{eq:NtotalSS} then yields
\begin{equation}
\frac{\gin}{\gout}=\frac{N_{\rm total}}{M_{k_G}}
\simeq110\quad\text{or}\quad310,
  \label{eq:exp_ratio}
\end{equation}
for the two pump powers with the temperature-renormalized rates given by
\begin{align}
 \goutbbar&=\frac{\gamma_{\rm out}^{\rm eff}}{k_G(1-q)\gammath},\notag\\
 \ginbbar&=\frac{\gammain}{k_G^2(1-q)\gammath}
 =\frac{\goutbbar}{k_G}\frac{\gin}{\gout}.
 \label{eq:numeric_rates}
\end{align}

\subsection{\texorpdfstring{Order-of-magnitude comparison}{Order-of-magnitude comparison}}

The coefficient $\gammath$ is not independently fixed by the experimental values above. To test whether the model reproduces the observed scale, we adopt a common value $\gammath=5\times10^6\;\mathrm{s}^{-1}$ for both pump powers. Since $\gammath$ represents a transition coefficient between individual modes, its inverse does not directly correspond to an experimentally measured thermalization or collision time. Using the remaining parameters specified above, the discrete balance equations~\eqref{eq:discrete-balance} yield
\begin{equation}
f_0^{(3\mathrm D)}\simeq
\begin{cases}
0.0208,&P_{\rm pump}=1.6\;\mathrm{mW},\\
0.0137,&P_{\rm pump}=8.8\;\mathrm{mW}.
\end{cases}
\label{eq:exp-fraction}
\end{equation}
Here, we use the exact degeneracies and truncate the spectrum at  $k_{\max}=k_G+8000$; increasing the cutoff to $k_G+12000$ changes either fraction by less than $10^{-8}$. The populations above the pumped shell are retained in the particle-number balance.

Both fractions are of order $10^{-2}$, consistent with the experimental observations. While this demonstrates phenomenological consistency at the percent-level scale, it constitutes neither an independent prediction nor quantitative agreement within experimental uncertainty. In particular, this choice of parameters does not reproduce the weak increase in the reported maximum fraction with pump power. Furthermore, the calculated values are sensitive to the assumed injection energy, bath temperature, and relaxation coefficient; for instance,  varying $\gammath^{-1}$ from 180 to 220 ns at fixed $T$ and $k_G$ shifts the resulting fractions from 0.0338 to 0.0116 at 1.6 mW and from 0.0235 to 0.00713 at 8.8 mW. Independent constraints on these effective parameters will therefore be required for a more rigorous quantitative test.

\subsection{Physical origin of the suppression}

The discrete results indicate that both particle loss and redistribution among low-lying excited states govern the ground-state fraction, with their relative contributions depending on the microscopic parameters. The present effective description can accommodate the small observed fraction; however, it does not disentangle the individual contributions  of finite lifetime, two-body loss, and incomplete thermal relaxation. Consequently, it neither establishes nor supports a quantitative prediction regarding variations in the experimental loss rate. 

\section{Conclusion}
BE condensation is a genuinely quantum-statistical phase transition in  that it occurs in the absence of interparticle interactions. This is because Bose symmetry dramatically reduces the number of available low-energy states, favoring the macroscopic occupation of the lowest single-particle state.
In thermal equilibrium, excited states can accommodate only a limited number of bosons, which decreases with decreasing temperature. Below a critical temperature, the maximum thermal capacity falls below the total number of bosons, forcing the remaining bosons to condense in the lowest single-particle state.

The situation dramatically changes for excitons whose ground state is the vacuum. A laser field creates excitons whose energy relaxes via thermal contact with the host semiconductor. We have formulated thermal relaxation so as to be consistent with detailed balance and satisfy the Kubo-Martin-Schwinger condition. In contrast to thermal equilibrium, the system exhibits a continuous downstream flow of bosons, whose steady-state distribution is governed not by the equilibrium BE distribution, but by the interplay among pumping, loss and thermal relaxation, rendering the resulting BE condensate    genuinely out of equilibrium. 

The ground-state fraction in the finite system varies smoothly across the parameter ranges studied, which neither establishes nor rules out a phase transition in every limiting procedure. From an experimental standpoint, it is worth noting that the total particle number is controlled by the pump-to-loss ratio rather than the chemical potential.  Our driven-dissipative theory enables a discrete calculation of the ground-state fraction in three dimensions. Using effective loss rates, an assumed injection energy, and a shared phenomenological relaxation coefficient, it yields percent-level fractions that agree in order of magnitude with the experiment by Morita et al. \cite{Morita2022}. While this comparison demonstrates that the model can capture the observed scale, achieving a predictive quantitative description will require  independent constraints on the effective parameters as well as a more explicit treatment of loss and relaxation.

\section*{Acknowledgments}
Fruitful discussions with Yusuke Morita,  Kosuke Yoshioka and Makoto Kuwata-Gonokami are gratefully acknowledged.
This work was supported by JSPS KAKENHI Grant No. 25K07178 and KAKENHI Grant No. JP22H01152 from the Japan Society for the Promotion of Science and the CREST program ``Quantum Frontiers" (Grant No. JPMJCR23I1) by the Japan Science and Technology Agency.
M.U. was supported by the RIKEN TRIP initiative.

\appendix

\section{Derivation of the mean-occupation rate equation}
\label{app:mean-occupation-derivation}

Here we derive Eq.~\eqref{eq:NumberEqExact} from the configuration-space
master equation~\eqref{eq:MasterP}.  We focus on a single mode $i=(k,\alpha)$ and define $\mN_i :=\langle N_i\rangle$ and 
\begin{equation}
\langle F(\vec N)\rangle := \sum_{\vec N}F(\vec N)P(\vec N).
\label{appendix1}
\end{equation}
We set $P(\vec N)=0$ if any component of $\vec N$ is negative, so that we can perform a change of variables in the particle-number configuration without introducing  boundary terms.

\subsection{Pump terms}
We first evaluate the two pump processes.  
Let $g:=(k_G,\alpha)$ denote a pumped mode.  Multiplying $N_i$ to the first term on the right-hand side of Eq.~(\ref{gain_A}) and summing over $\vec{N}$, we have
\begin{align}
 \sum_{\vec{N},\alpha}N_iN_gP(\vec N-\vec e_g)
=\sum_{\vec{N},\alpha} (N_i+\delta_{ig})(N_g+1) P({\vec N}),
 \label{eq:creation-shift}
\end{align}
where $\delta_{i,g}$ is the Kronecker delta on the composite mode
indices.
Similarly, the second term on the right-hand side of Eq.~(\ref{gain_A}) leads to
\begin{align}
 \sum_{\vec{N},\alpha}N_i(N_g+1)P(\vec N).
 \label{eq:creation-shift2}
\end{align}
Subtracting this from Eq.~(\ref{eq:creation-shift}), we obtain
\begin{equation}
\frac{d\mathcal{N}_i}{dt}\biggr|^\mathrm{up}_\mathrm{pump}
=\gamma_\mathrm{in}\sum_{\alpha=1}^{M_{k_G}}\delta_{i,g}\langle N_g+1\rangle
\label{pump_up2}
\end{equation}
The contributions from the first and second terms on the right-hand side of Eq.~(\ref{gain_B}) are
\begin{eqnarray}
&&\sum_{{\vec N},\alpha}N_i(N_g+1)P({\vec N}+{\vec e}_g)
=\sum_{{\vec N},\alpha}(N_i-\delta_{ig})N_gP({\vec N})
\nonumber \\
&&\sum_{{\vec N},\alpha}N_iN_gP({\vec N})
\nonumber
\end{eqnarray}
Subtracting the second line from the first, we obtain
\begin{equation}
\frac{d\mathcal{N}_i}{dt}\biggr|^\mathrm{down}_\mathrm{pump}
=-\gamma_\mathrm{in}\sum_{\alpha=1}^{M_{k_G}}\delta_{i,g}\langle N_g\rangle
\label{pump_down2}
\end{equation}
Summing Eqs.~(\ref{pump_up2}) and (\ref{pump_down2}), we obtain
\begin{equation}
 \left.\frac{d\mN_{k\alpha}}{dt}\right|_{\rm pump}
 =\delta_{k,k_G}\gamma_{\rm in}.
 \label{eq:gain-mean-result}
\end{equation}
Thus, the occupation-dependent pieces cancel each other, and each mode at the level $k_G$ is pumped at the constant rate of $\gamma_{\rm in}$.

\subsection{Loss terms} 
We next evaluate the two loss processes in Eq.~(\ref{loss_A}).
Multiplying the first and second terms on the right-hand side of Eq.~(\ref{loss_A}) by $N_i$ and summing over $\vec{N}$, we obtain
\begin{eqnarray}
&&\sum_{\vec{N},p}N_i(N_p+1)P({\vec N}+{\vec e}_p)=\sum_{\vec{N},p}(N_i-\delta_{ip})N_p P(\vec{N}),  \nonumber \\
&&\sum_{\vec{N},p}N_iN_pP({\vec N}). \nonumber  
\end{eqnarray}
Subtracting the second line from the first leads to
\begin{equation}
 \left.\frac{d\mN_{k\alpha}}{dt}\right|_{\rm loss}
 =-\gamma_{\rm out}\mN_{k\alpha}.
 \label{eq:loss-mean-result}
\end{equation}

\subsection{Thermal relaxation}
The contribution of thermal relaxation to the rate equation of mode $i$ is given by
\begin{equation}
\frac{d\mathcal{N}_i}{dt}\biggr|_{\rm R}=\sum_{\vec N} D_{\rm R}[P] N_i, 
\label{TR1}
\end{equation}
where $D_{\rm R}[P]$ is given in Eq.~(\ref{TR_dissipator}). 
To evaluate this, we consider the contribution from the first term on the right-hand side of Eq.~(\ref{TR_dissipator}) by focusing on a pair of distinct modes $p$ and $q$ and set $F({\vec N})=N_i N_p(N_q+1)$ in Eq. (\ref{appendix1}):
\begin{equation}
\sum_{\vec N}N_i (N_p+1)N_q P({\vec N}+{\vec e}_p-{\vec e}_q)
\label{appendix2}
\end{equation}
We make a change of variables ${\vec N}\rightarrow\vec N-\vec e_p+\vec e_q$, obtaining
\begin{align}
 &\sum_{\vec N}(N_i-\delta_{ip}+\delta_{iq})N_p(N_q+1)
 P(\vec N)
 \label{eq:thermal-shift}
\end{align}
Similarly, the second term in Eq.~(\ref{TR_dissipator}) gives 
\begin{align}
 &\sum_{\vec N}N_iN_p(N_q+1) P(\vec N)
 \label{eq:thermal-shift2}
\end{align}
Subtracting this from Eq.~(\ref{eq:thermal-shift}) leads to 
\begin{equation}
 \Gamma_{p\to q}(\delta_{iq}-\delta_{ip})
 \left\langle N_p(N_q+1)\right\rangle .
 \label{eq:thermal-pair-contribution}
\end{equation}
Only transitions entering or leaving mode $i$ survive.  Restoring the
original mode indices and retaining them on the rates, we obtain
\begin{align}
 \left.\frac{d\mN_{k\alpha}}{dt}\right|_R
 &=\sum_{k'(\ne k)}\sum_{\alpha'=1}^{M_{k'}}
 \Gamma_{k'\alpha'\to k\alpha}
 \left\langle (N_{k\alpha}+1)N_{k'\alpha'}\right\rangle\notag\\
 &\quad-\sum_{k'(\ne k)}\sum_{\alpha'=1}^{M_{k'}}
 \Gamma_{k\alpha\to k'\alpha'}
 \left\langle N_{k\alpha}(N_{k'\alpha'}+1)\right\rangle .
 \label{eq:thermal-mean-derivation}
\end{align}

Adding Eqs.~\eqref{eq:thermal-mean-derivation},
\eqref{eq:gain-mean-result}, and \eqref{eq:loss-mean-result} and 
summing over $\alpha=1,\ldots,M_k$ gives Eq.~\eqref{eq:NumberEqExact}.

\bibliography{refs}
\end{document}